\documentstyle[aps,prl,preprint]{revtex}

\begin{document}
\textheight 8.5in
\textwidth  6.in

\def\la{\mathrel{\mathpalette\fun <}}
\def\ga{\mathrel{\mathpalette\fun >}}
\def\fun#1#2{\lower3.6pt\vbox{\baselineskip0pt\lineskip.9pt
        \ialign{$\mathsurround=0pt#1\hfill##\hfil$\crcr#2\crcr\sim\crcr}}}
\def\l{\lambda}
\def\n{n(R,t)}
\def\f{\phi}
\vspace*{-62pt}
\begin{flushright}
DART-HEP-95/03\\
July 1995
\end{flushright}

\vspace{0.75in}
\centerline{\bf THERMAL MIXING OF PHASES:}
\centerline{\bf NUMERICAL AND ANALYTICAL STUDIES
\footnote{Invited talk given at the ``3\'eme Colloque Cosmologie'', Paris,
June 7--9, 1995.}}

\vskip 1.cm
\centerline{Marcelo Gleiser\footnote{NSF Presidential Faculty Fellow}}

\vskip .5 cm
\centerline{Department of Physics and Astronomy}
\centerline{Dartmouth College}
\centerline{Hanover, NH 03755, USA}

\def\mpl{{m_{Pl}}}
\def\x{{\bf x}}
\def\p{\phi}
\def\F{\Phi}
\def\s{\sigma}
\def\a{\alpha}
\def\d{\delta}
\def\t{\tau}
\def\r{\rho}
\def\beq{\begin{equation}}
\def\eeq{\end{equation}}
\def\ba{\begin{eqnarray}}
\def\ea{\end{eqnarray}}
\def\re#1{^{\ref{#1}}}

\vskip .5 cm
\centerline{\bf Abstract}
\begin{quote}
\baselineskip 12pt
The dynamics of phase transitions plays a crucial r\^ole in the so-called
interface between high energy particle physics and cosmology. Many of the
interesting results generated during the last fifteen years or so rely
on simplified assumptions concerning the complex mechanisms typical of
nonequilibrium field theories. In particular, whenever first order
phase transitions are invoked, the metastable background is assumed to
be sufficiently smooth to justify the use of homogeneous nucleation
theory in the computation of nucleation rates of critical bubbles.
In this talk I present the results of numerical simulations which were
designed to quantify ``smoothness''; that is, how the contribution from
nonperturbative
subcritical fluctuations may spoil the homogeneity assumption of nucleation
theory. I then show how the numerical results can be understood
{\it quantitatively} in terms of a simple analytical
model of subcritical thermal
fluctuations. Encouraged by the success of the model in matching the
numerical results, I apply it to the
standard model electroweak phase transition.

\end{quote}

\baselineskip 14pt
\def\beq{\begin{equation}}
\def\eeq{\end{equation}}
\def\ba{\begin{eqnarray}}
\def\ea{\end{eqnarray}}
\def\re#1{[{\ref{#1}]}}

\def\mpl{{m_{Pl}}}
\def\x{{\bf x}}
\def\p{\phi}
\def\F{\Phi}
\def\s{\sigma}
\def\a{\alpha}
\def\d{\delta}
\def\t{\tau}
\def\r{\rho}

\vspace{16pt}
\section{Introduction: Homogeneous Nucleation and its Assumptions}

The fact that the gauge symmetries describing particle interactions
can be restored at high enough temperatures has led, during the past
15 years or so, to an active research program on the possible
implications that this symmetry restoration might have had to the
physics of the very early Universe. One of the most interesting and
popular possibilities is that during its expansion the Universe
underwent a series of phase transitions, as some higher symmetry group
was successively broken into products of smaller groups, up to the
present standard model described by the product $SU(3)_C\otimes
SU(2)_L \otimes U(1)_Y$. Most models of inflation and the formation of
topological (and nontopological) defects are well-known consequences
of taking the existence of cosmological phase transitions seriously
\cite{KT}.

One, but certainly not the only,
motivation of the works  addressed in this talk
comes from the possibility
that the baryon asymmetry of the Universe could have been dynamically
generated during a first order electroweak phase transition \cite{EW}.
As is by now clear, a realistic calculation of the net baryon number
produced during the transition is a formidable challenge. We probably
must invoke physics beyond the standard model (an exciting prospect
for most people) \cite{FS}, push perturbation theory to its limits
(and beyond, due to the nonperturbative nature of magnetic plasma
masses that regulate the perturbative expansion in the symmetric
phase), and we must deal with nonequilibrium aspects of the phase
transition. Here I will focus on the latter problem, as it seems to
me to be the least discussed of the pillars on which most baryon
number calculations are built upon. To be more specific, it is possible
to separate the nonequilibrium aspects of the phase transition into two
main subdivisions. If the transition proceeds by bubble nucleation, we
can study the propagation of bubbles in the hot plasma and the
transport properties of particles through the bubble wall. A
considerable amount of work has been devoted to this issue, and the
reader can consult the works of Ref. \cite{BW} for details. These
works assume that homogeneous nucleation theory is adequate to
investigate the evolution of the phase transition, at least for the
range of parameters of interest in the particular model being used to
generate the baryon asymmetry. This brings us to the second important
aspect of the nonequilibrium dynamics of first order phase
transitions, namely the validity of homogeneous nucleation theory to
describe the approach to equilibrium. This is the issue addressed in
this talk.

Nucleation theory is a well-studied, but far from exhausted, subject.
Since the pioneering work of Becker and D\"oring on the nucleation of
droplets in supercooled vapor \cite{BD}, the study of first order
phase transitions has been of interest to investigators in several
fields, from meteorology and materials science to quantum field theory
and cosmology. Phenomenological field theories were developed by Cahn
and Hilliard and by Langer \cite{CH,LANGER} in the context of
coarse-grained time-dependent Ginzburg-Landau models, in which an
expression for the decay rate per unit volume was obtained by assuming
a steady-state probability current flowing through the saddle-point of
the free-energy functional \cite{LANGER,DOMB}. The application of metastable
decay to quantum field theory was initiated by Voloshin, Kobzarev, and
Okun \cite{VKO}, and soon after put onto firmer theoretical ground by
Coleman and Callan \cite{CC}. The generalization of these results for
finite temperature field theory was first studied by Linde
\cite{LINDE}, and has been the focus of much recent attention
\cite{FINITETDECAY}.

The crucial ingredient in the evaluation of the decay rate is the
computation of the imaginary part of the free energy. As shown by
Langer \cite{LANGER}, the decay rate ${\cal R}$ is proportional to the
imaginary part of the free energy ${\cal F}$,
\beq
{\cal R} = {{\mid E_-\mid }\over {\pi T}}{\rm Im} {\cal F} ~,
\eeq
where $E_-$ is the negative eigenvalue related to metastability, which
depends on nonequilibrium aspects of the dynamics, such as the
growth rate of the critical bubble.
Since ${\cal F}= - T {\rm ln}
Z$, where $Z$ is the partition function, the computation for the rate
boils down to the evaluation of the partition function for the system
comprised of critical bubbles of the lower energy phase inside the metastable
phase.

If we imagine the space of all
possible field
configurations for a given model,
there will be different paths to go from the metastable
to the ground state. We can think of the two states as being separated
by a hill of a given ``height''.
The energy barrier for the decay is then related
to the height of this hill. At the top of the hill, only one direction
leads down to the ground state, the unstable direction. Fluctuations
about this direction will grow, with rate given by the negative
eigenvalue which appears in the above formula. All other directions are
positively curved, and fluctuations about them give rise to positive
eigenvalues which do not contribute to the decay rate.
The path which will cost less energy is the one which will
dominate the partition function, the so-called critical bubble or bounce.
It is simply the field configuration that interpolates between the two
stable points in the {\it energy landscape}, the metastable and ground state.
The energy barrier for the decay is the energy of this particular field
configuration.

For a dilute gas of bubbles only, the partition function for
several bubbles is given by \cite{AM,LANGER},
\ba
Z & \simeq & Z(\varphi_{f}) +Z(\varphi_{f}) \left[ \frac{Z(\varphi_{b})}
{Z(\varphi_{f})} \right] + Z( \varphi_{f}) \frac{1}{2 !} \left[
\frac{Z(\varphi_{b})}{Z(\varphi_{f})} \right]^{2} + \ldots
\nonumber \\
& \simeq & Z(\varphi_{f}) \exp \left[ \frac{Z(\varphi_{b})}{Z(\varphi_{f})}
\right]
\: ,
\label{e:Zmany}
\ea
where $\varphi_{f}$ is the metastable vacuum field configuration and
$\varphi_{b}$ is the bubble configuration, the bounce solution to the
$O(3)$-symmetric Euclidean equation of motion. We must evaluate the
partition functions above. This is done by the saddle-point method,
expanding the scalar field $\phi({\bf x},\tau)$,
such that $\phi({\bf x},\tau) \rightarrow \varphi_{f}
+\zeta(\bf{x} ,\tau)$ for $Z(\varphi_{f})$, and $ \phi({\bf x},\tau)
\rightarrow \varphi_{b} (\bf{x}) +\eta (\bf{x} ,\tau)$ for
$Z(\varphi_{b})$, where $\zeta(\bf{x} ,\tau)$ and $\eta (\bf{x} ,\tau)$
are small fluctuations about equilibrium.

It is crucial to note that the saddle-point, or Gaussian, method only
gives good results if indeed the fluctuations about
equilibrium are sufficiently small that nonlinear terms in the fields
can be neglected. Even though the method sums over all amplitude fluctuations,
it does so by assuming that the functional integral
is well approximated by truncating
the expansion of the action to second order. The efficiency of the method
relies on the fact that higher amplitudes will be suppressed fast enough
that their contribution to the partition function will be negligible. One
can visualize this by comparing a sharp parabolic curve with a flatter
one with minimum at $x_0$,
and investigating when $\int dx e^{-f(x)}$ will be well approximated
by writing $f(x)\simeq f(x_0) + {1\over 2}(x-x_0)^2f^{\prime\prime}(x_0)$.
For a sharp curve, larger amplitude fluctuations will be strongly
suppressed and thus give a negligible contribution to
the integral over all amplitudes. Clearly, this will not be the case for
flatter curves.

Skipping details \cite{FINITETDECAY}, using the saddle-point method
one obtains for the ratio of partition functions,
$\frac{Z(\varphi_{b})}{Z(\varphi_{f})}$,
\begin{equation}
\frac{Z(\varphi_{b})}{Z(\varphi_{f})} \stackrel{saddle-point}{\simeq}
\left[ \frac{\det ( -\Box_{E} + V''(\varphi_{b}))_{\beta}}
{ \det ( -\Box_{E} + V''(\varphi_{f}))_{\beta}} \right]^{-\frac{1}{2}}
e^{-\Delta S} \: ,
\label{ratiodet}
\end{equation}
where $[ \det (M)_{\beta}]^{- \frac{1}{2}} \equiv \int D \eta \exp
\left\{ - \int_{0}^{\beta} d \tau \int d^{3} x \frac{1}{2} \eta [M]
\eta \right\}$ and $\Delta S = S_{E}(\varphi_{b})-S_{E}(\varphi_{f})$
is the difference between the Euclidean actions for the field
configurations $\varphi_{b}$ and $\varphi_{f}$. [Note that
$S_{E}(\varphi)$, and hence $\Delta S$, does not include any
temperature corrections. It would if one had summed over other fields
coupled to $\varphi$.] Thus, the free energy of the system is,
\beq\label{energy}
{\cal F} = - T \left[ \frac{ \det ( -\Box_{E} + V''(\varphi_{b}))_{\beta}}
{ \det ( -\Box_{E} + V''(\varphi_{f}))_{\beta}} \right]^{-\frac{1}{2}}
e^{-\Delta S} \: .
\eeq

Let me stress again the assumptions that go into computing the free energy.
First, that
the partition function is given by Eq. \ref{e:Zmany} within the
dilute gas approximation, and second, that the partition function is
evaluated approximately by
assuming {\it small} fluctuations about the homogeneous
metastable state $\varphi_{f}$. It is clear that for situations in
which there are large amplitude fluctuations about the metastable
equilibrium state the above formula must break down. Thus the
breakdown of the expression for the rate is intimately connected with
the question of how well-localized the system is about the metastable
state as the temperature drops below the critical temperature $T_c$.
Homogeneous nucleation, as its name already states, is only accurate when
the metastable state is sufficiently homogeneous. In the presence of
inhomogeneities, there is no reason to expect that the decay rate formula
will apply. The question then is to quantify {\it when} does it
breakdown.

In order to study the properties of the metastable background in the
presence of thermal fluctuations, I have recently investigated the
nonequilibrium dynamics of a (2+1)-dimensional model of a scalar field
coupled to a thermal bath \cite{MG}. The nonlinear interactions were
chosen to reflect the gross properties of the electroweak effective
potential, although the model only deals with a real scalar field. These
results were then extended to a (3+1)-dimensional simulation, in work done in
collaboration with Julian Borrill \cite{BG}. The results for two and three
dimensions were qualitatively similar, showing how the amount of
phase mixing is related to the parameters of the model. In particular,
for the three-dimensional study we varied the strength of the scalar field's
quartic coupling, which for the electroweak model is related to the Higgs boson
mass. In the next Section I review the results of this work. In Section 3
I present an improved treatment of the kinetics of subcritical bubbles which
I then use to understand the results of the numerical simulations.
The agreement between numerical experiment and analytical model is extremely
satisfactory. The simplicity of the kinetic approach based on the
dynamics of subcritical thermal fluctuations makes it a useful tool
in studying the properties of the thermal background. I then briefly
apply the subcritical bubbles method to the simple 1-loop electroweak
potential, showing that for Higgs masses above $60$ GeV, the amount of
phase mixing is such that homogeneous nucleation should not be used to
study the dynamics of the transition. I conclude in Section 4 with
a brief overview of the results and possible future work.

\section{Numerical Simulations of Thermal Phase Mixing}

The homogeneous part of the free energy density is written as
\beq
U(\p ,T)={a\over 2}\left (T^2-T_2^2\right )\p^2-
{{\alpha}\over 3}T\p^3 +{{\l}\over 4}
\p^4 \:.
\label{e:freen}
\eeq
This choice intentionally resembles the electroweak effective
potential to some order in perturbation theory, although here $\p({\bf
x},t)$ is a real scalar field, as opposed to the magnitude of the
Higgs field. The goal is to explore the possible dynamics of a model
described by the above free-energy density, generalizing the results
obtained in Ref. \cite{MG} to (3+1)-dimensions. The analogy with the
electroweak model is suggestive but not quantitative.

Introducing dimensionless variables ${\tilde x} = a^{{1}\over{2}} T_2
x$, ${\tilde t} = a^{{1}\over{2}} T_2 t$, $X = a^{-{{1}\over{4}}}
T_2^{-1} \p$, and $\theta = T/T_2$, the Hamiltonian is,
\beq
{{H[X]}\over {\theta}}={1\over {\theta}}\int d^2{\tilde x}\left [
{1\over 2}\mid {\tilde \bigtriangledown} X\mid^2 +
{1\over 2}\left (\theta^2 -1
\right )X^2 -{{
{\tilde \alpha}}\over 3}\theta X^3+{{{\tilde \l}}\over 4}X^4\right ] \:,
\label{e:hamilton}
\eeq
where ${\tilde \alpha} = a^{-{{3}\over{4}}} \alpha$, and ${\tilde
\lambda} = a^{-{{1}\over{2}}} \lambda$ (henceforth we drop the
tildes). For temperatures above $\theta_1=(1-\a^2/4\l)^{-{{1}\over
2}}$ there is only one minimum at $X=0$. At $\theta=\theta_1$ an
inflection point appears at $X_{\rm inf}=\alpha \theta_1 /2\lambda$.
Below $\theta_1$ the inflection point separates into a maximum and a
minimum given by $X_{\pm}={{\a\theta}\over {2\l}}\left [ 1\pm
\sqrt{1-4\l\left (1-1/\theta^2\right )/\a^2}\right ]$. At the critical
temperature $\theta_c=(1-2\a^2/9\l)^{-{{1}\over 2}}$ the two minima,
at $X_0=0$ and $X_+$ are degenerate. Below $\theta_c$ the minimum at
$X_+$ becomes the global minimum and the $X_0$-phase becomes
metastable. Finally, at $\theta=1$ the barrier between the two phases
at $X_-$ disappears.

The coupling with the thermal bath will be modelled by a Markovian
Langevin equation which, in terms of the dimensionless variables
defined above, is
\beq\label{e:langevin}
{\partial^2X\over\partial t^2} = \nabla^2 X - \eta {\partial X\over
\partial t} - {\partial U(X,\theta) \over \partial X} + \xi(x,t)~~,
\eeq
where $\eta$ is the dimensionless viscosity coefficient, and $\xi$ the
dimensionless stochastic noise with vanishing mean, related to $\eta$
by the fluctuation-dissipation theorem,
\beq\label{e:fluc-diss}
\langle \xi({\bf x},t)\xi( {\bf x'},t')\rangle =
  2 \eta \theta \delta(t-t')\delta^3({\bf x} - {\bf x'})~~.
\eeq
A few comments are in order concerning our choice of equation. It is
clear that we are assuming that $X({\bf x},t)$ represents the
long-wavelength modes of the scalar field. Whenever one discretizes a
continuum system there is an implicit coarse-graining scale built in.
We encapsulate information about the shorter-wavelength modes, which
have faster relaxation time-scales, in the dissipation and noise
terms. In principle it should be possible to derive an effective
Langevin-like equation for the slow modes by integrating out the fast
modes from the effective action. This is a complicated problem, and
progress has been slow. Recent work indicates that one should expect
departures from the Langevin equation written above \cite{LANGEVIN},
although details are sensitive to the particular model one starts
with. For example, the noise may be colored (with more complicated
correlation functions) and the coupling to the bath may be
multiplicative, as opposed to the additive coupling chosen above.
Here, we will adopt the above equation as a first step. We do not
expect that the nature of the noise will change the final equilibrium
properties of the system, but mostly the relevant relaxation
time-scales. Furthermore, Gleiser and Ramos showed that for high enough
temperatures the noise does become white \cite{GR}.
Since the physical results here are related to the final
equilibrium state of the system, we believe that they will not be
affected by more complicated representations of the coupling of the
field to the thermal bath. However, a more thorough examination of this
question deserves further study.

A related topic is the choice of coarse-graining scale, which is
embedded in the lattice spacing used in the simulations. It is
well-known that any classical field theory in more than one spatial
dimension is ultra-violet divergent, and that the lattice spacing
serves as an ultra-violet cutoff. This being the case, one should be
careful when mapping from the lattice to the continuum theory. If one
is to probe physics at shorter wavelengths, renormalization
counterterms should be included in the lattice formulation so that a
proper continuum limit is obtained on the lattice within the validity of
perturbation theory. This point has been
emphasized in Ref. \cite{AG}, where a (2+1)-dimensional study of
nucleation was performed for a temperature-independent potential.
Renormalization counterterms (of order $\theta ~{\rm ln}\,\delta x$
for lattice spacing $\delta x$) for a particular renormalization
prescription were obtained, and the results shown to be lattice-space
independent.

Here, due to the temperature dependence of the potential, the
renormalization prescription of Ref. \cite{AG} does not work.
Instead, we will use $\delta x = 1$ throughout this work. It turns out
that for all cases studied the mean-field correlation length
$\xi^{-2}=V^{\prime \prime} (X_0,\theta_c)$ is sufficiently larger
than unity to justify this choice. Modes with shorter wavelengths
are coupled through the noise into the dynamics of the longer
wavelength modes, as described by Eq. \ref{e:langevin} above.
But it should be stressed that different choices of lattice spacing
imply in different effective coarse-graining scales, and hence different
results. In other words, although phase mixing will always be present to
some degree, the
exact quantitative results will depend on the coarse-graining scale.

I will skip details of how we implemented this simulation in a parallel
machine, as well as the tests to make sure our results were independent of
simulation parameters such as lattice length, time and spatial steps, and
random number generator. The reader is referred to Ref. \cite{BG} for
information on these issues. I will now proceed by describing the numerical
experiments and our results.

As pointed out in the Introduction, the question of whether homogeneous
nucleation theory is trustworthy to describe a first order phase
transition boils down to how well localized in the
metastable state the system is, as the temperature drops below the critical
temperature. In the jargon of condensed matter physics, nucleation should work
for quenches deep into the metastable branch of the coexistence curve.
In order to address this
question, following the procedure of Ref. \cite{MG}, we will study the
behavior of the system at the critical temperature, when the two
minima are degenerate. The reason for this choice follows naturally
from the fact that we are interested on the way by which the system
approaches equilibrium as the temperature drops below $T_c$. The
detailed dynamics will depend on the relative fraction of the total
volume occupied by each phase; if at $T_c$ the system is well
localized about the $X=0$ minimum, as the temperature drops the
transition may evolve by nucleation and subsequent percolation of
bubbles larger than a critical size.  If, on the other hand,
considerable phase-mixing occurs already at $T_c$, we expect the
transition to evolve by domain coarsening, with the domains of the
$X_+$ phase eventually permeating the whole volume.

Let us call the two phases the $0$-phase and the $+$-phase,
corresponding to the local equilibrium values $X=X_0=0$, and $X=X_+$,
respectively. We can quantify the phase distribution of the system as
it evolves according to Eq. \ref{e:langevin}, by measuring the
fraction of the total volume in each phase. This is done by simply
counting the total volume of the system at the left of the potential barrier's
maximum
height ({\it i.e.}, $X\leq X_-\equiv X_{\rm max}$),
corresponding to the $0$-phase.
Dividing by the total volume, we obtain the fraction of the system in
the $0$-phase, $f_0(t)$, such that
\beq
f_0(t) + f_+(t) = 1 ~~,
\eeq
where, of course, $f_+(t)$ corresponds to the fractional volume in the
$+$-phase. A further measure of any configuration is given by the
volume-averaged order parameter, $\langle X \rangle (t)= V^{-1}\int
dV~X(t)$. A localized configuration ($f_0^{\rm eq} > 0.5$) then
corresponds to $\langle X \rangle_{\rm eq} < X_{\rm max}$, and a fully
phase-mixed configuration ($f_0^{\rm eq} \simeq  0.5$) to $\langle X
\rangle_{\rm eq} = X_{\rm max}$, where the super(sub)-script `eq' refers
to final ensemble-averaged equilibrium values of $f_0(t)$ and
$\langle X \rangle (t)$, respectively. [Recall that in the presence of noise
we must take an ensemble average of the physical quantities of interest in
order to get smooth, sensible results. In practice, this amounts to changing
the seed of the random number generator for each run, and averaging the
results over many runs.]

We prepare the system so that initially it is well localized in the
$0$-phase, with $f_0(0)=1$ and $\langle X \rangle (0)=0$. These
initial conditions are clearly the most natural choice for the problem
at hand. If one has cosmology in mind, it is quite possible that as
the system slowly cools down (we are not interested in phase
transitions close to the Planck scale), fluctuations from the high
temperature phase $X=0$ to the $X_+$ phase are already occurring before
$T_c$ is reached. (In this case, our arguments are even stronger.)
However, we will adopt the best-case scenario for homogeneous
nucleation to work, in which the system managed to reach the $X=0$
phase homogeneously, so that the initial state is a thermal state with
mean at $X_0$.  If one has more concrete applications in mind, we can
assume that we quenched the system to its critical temperature, making
sure that the order parameter remains localized about the
high-temperature phase.  Since thermalization happens very fast in the
simulations, the exact point by point initial conditions should not be
important, and we can view the first few time steps as generating an
initial thermal distribution with $f_0(0) \sim 1$ and $\langle X
\rangle \sim 0$, so that the average kinetic energy per lattice point
satisfies the equipartition theorem, ${1\over N}E_k = {3\over 2}T$.
For simplicity we take $X=0, \dot X=0$ everywhere initially.

There are two parameters controlling the strength of the transition,
$\alpha$ and $\lambda$. In the previous (2+1)-dimensional work,
$\alpha$ was chosen to vary while $\lambda$ was kept fixed. It is
really immaterial which parameter is held fixed, or if both are made
to vary, but in order to keep closer to the spirit of the electroweak
model we will fix $\alpha$ and let $\lambda$ vary. As is well-known,
$\lambda$ is related to the Higgs mass, while $\alpha$ is related to
the gauge-boson masses \cite{EW}. The connection with the
electroweak model is straightforward. If we consider as an example the
unimproved one-loop approximation, the effective potential is
\cite{EW},
\beq
\label{e:VEW}
V_{{\rm EW}}(\p,T)=D\left (T^2-T_2^2 \right )\p^2-ET\p^3+{1\over
4}\l_T\p^4,
\eeq
where $D$ and $E$ are given by
$D=\left[6(M_W/\s)^2+3(M_Z/\s)^2+6(M_T/\s)^2\right ]/24\simeq 0.17$
and $E=\left [6(M_W/\s)^3+3(M_Z/\s)^3\right ]/12\pi\simeq 0.097$, for
$M_W=80.6$ GeV, $M_Z=91.2$ GeV, $M_T=174$ GeV \cite{MT}, and $\s=246$
GeV. $T_2$ is given by,
\beq
\label{eq:T2}
T_2=\sqrt{(M_H^2-8B\s^2)/4D}\ ,
\eeq
where the physical Higgs mass is given in terms of the 1-loop
corrected $\l$ as $M_H^2=\left (2\l+12B\right) \s^2$, with $B=\left
(6M_W^4+3M_Z^4-12M_T^4\right )/64\pi^2\s^4$, and the
temperature-corrected Higgs self-coupling is,
\beq
\l_T=\l-{1\over {16\pi^2}}\left [
\sum_Bg_B\left ({{M_B}\over {\s}}\right )^4
{\rm ln}\left (M_B^2/c_BT^2\right )-\sum_Fg_F\left ({{M_F}\over {\s}}
\right )^4{\rm ln}\left (M_F^2/c_FT^2\right )\right]
\eeq
where the sum is performed over bosons and fermions (in our case only
the top quark) with their respective degrees of freedom $g_{B(F)}$,
and ${\rm ln}c_B=5.41$ and ${\rm ln}c_F=2.64$.

Thus, the correspondence with our (dimensionless) parameters is
\beq\label{e:corresp}
\alpha = {{3E}\over {(2D)^{3\over 4}}}=0.065~,~{\rm and}~~~\l = {{\l_T}\over
{(2D)^{1\over 2}}}=1.72 \l_T~~.
\eeq
Once this is established, the numerical experiment proceeds as
follows: i) Choose $\alpha=0.065$; ii) Prepare the system in the initial
state described above, and measure the value of $f_0(t)$ and
$\langle X\rangle (t)$ for several values of $\lambda$, as the system
evolves according to Eq. \ref{e:langevin}. I can now present the
results of the simulations.

Based on the above discussion, we choose lattice length $L=48$,
lattice spacing $\delta x = 1$,
time step $\delta t = 0.1$, and $\alpha = 0.065$ in all simulations.
[For details
as to why this choice of parameters is sensible, see Ref. \cite{BG}.] The
experiment then consists in measuring the fraction of the volume in
the $0$-phase as a function of time for several values of $\lambda$.

In Fig. 1 we show the evolution of the ensemble-averaged fraction
$f_0(t)$ for several values of $\l$. It is clear that for small enough
values of $\lambda$ the system remains well-localized in the $0$-phase
with $f_0^{\rm eq}\sim 1$, while for larger values the two phases
become completely mixed, with $f_0^{\rm eq}\rightarrow 0.5$.
Remarkably, the transition region between the two regimes is quite
narrow, centered around $\l \simeq 0.025$. This can be seen from Fig.
2 where we show $f_0(t)$ for $\l = 0.024,~0.025$, and $0.026$. [The
curves are noisier due to the fact that we must run for longer times in
order to approach the equilibrium values of $f_0(t)$, being thus
constrained to perform an ensemble average with fewer runs.]


\vspace{2.cm}

{\bf Figure 1:} The approach to equilibrium for several values of $\l$. \\

\vspace{2.cm}

{\bf Figure 2}: Fitting $f_0(\theta_c)$ to a power law at large times for
$\l=0.025$.\\

\vspace{2.cm}

Note that
for $\l=0.026$, $f_0^{\rm eq}\simeq 0.5$, while for $\l = 0.024$,
$f_0^{\rm eq} \simeq 0.72$. There is a pronounced change in the
behavior of the system for $\l \simeq 0.025$. Furthermore, we find
that the numerical curves can be fitted at all times by a stretched
exponential,
\beq
\label{e:expfit}
f_0(t) = \left (1-f_0^{\rm eq}\right ){\rm exp}
\left [-(t/\tau_{\rm eq})^{\s}\right ] + f_0^{\rm eq}~~,
\eeq
where $f_0^{\rm eq}$ is the final equilibrium fraction and $\tau_{\rm
eq}$ is the equilibration time-scale. In Table 1 we list $\s$ and
$\tau_{\rm eq}$ for several values of $\lambda$. Note that for $\l =
0.025$ the fit is obtained at late times by a power law, (smooth curve
in Fig. 5)
\beq
\label{e:powerlaw}
f_0(t)\mid_{\l=0.025}~~  \propto ~~t^{-k}~~,
\eeq
with $k = 0.10(\pm 0.02)$. This slowing down of the approach to
equilibrium is typical of systems in the neighborhood of a second
order phase transition, being known as `critical slowing down'
\cite{SLOW}.  This behavior is suggestive of a ``phase
transition'' between two possible regimes for the system, one in which
the system is well-localized in the $0$-phase, and the other in which
there is a complete mixing between the two phases.  Let us call these
two regimes the `smooth' and the `mixed' regimes, respectively. [In
Refs. \cite{MG} and \cite{BG} I used the terms ``strong'' and ``weak'',
respectively. The change is not to induce confusion, but simply to
emphasize that the difference between the two regimes is in the amount
of mixing present in the background. Strong and weak refer to the
strength of the associated phase transition, although I consider the term
weak first order transition a misnomer; the question is nucleation vs.
spinodal decomposition and not weak vs. strong.]
Before
we explore this idea any further, I note that the final
equilibrium fractions are insensitive to the viscosity parameter $\eta$,
which reflects the coupling of the system to the thermal bath. For details,
I refer the reader to Ref. \cite{BG}.
This is precisely what one
expects, as the coupling to the bath should not influence the final
equilibrium properties of the system, but only how fast it
equilibrates.

\vspace{2.cm}

{\bf Table 1}:  The values of the equilibration time-scales
and the exponents for
the exponential fit of Eq. \ref{e:expfit}
for several values of $\l $.
Uncertainties are
in the last digit.

\vspace{2.cm}

Armed with these results, and invoking also the results in $(2+1)$-dimensions
\cite{MG}, we define the equilibrium fractional population difference
\beq\label{e:Delta}
 \Delta F_{\rm EQ}\left (\theta_c \right ) = f_0^{\rm eq} - f_+^{\rm eq}~~.
\eeq
In Fig. 3 we show the behavior of $\Delta F_{\rm EQ}$ as a function of
$\l$.  There is a clear qualitative analogy between the behavior of
$\Delta F_{\rm EQ}$ as a function of $\l$ and the behavior of the
magnetization as a function of temperature in Ising models.  Here, the
order parameter is the equilibrium fractional population difference
and the control parameter is $\l$. $\l_c\simeq 0.025$ is the critical
value for the parameter $\l$, which determines the degree of mixing
of the system at $T_c$.

\vspace{2.cm}
{\bf Figure 3}:  The fractional equilibrium population difference $\Delta
F_{{\rm EQ}}$ as a function of $\l$. \\

\vspace{2.cm}

We stress that the idea here is to probe the assumption of
localization within the $0$-phase as the system cools to $T_c$. Our
results show that if the time-scales for cooling are slower than the
equilibration time-scales of the system, for $\l > \l_c$ there will be
considerable phase mixing before the temperature drops below $T_c$.
This result can be made quite transparent by comparing the equilibrium
value of the volume-averaged field $\langle X\rangle_{\rm eq}$, and
the location of both the inflection point and the maximum of the
potential with varying $\l$. As can be seen from Fig. 4, the narrow
transition region is clearly delimited by
\beq
X_{\rm inf}~~ <~~ \langle X\rangle_{\rm eq} ~~< X_{\rm max}~~,
\eeq
where $X_{\rm inf}$ and $X_{\rm max}$ are the inflection point and the
maximum of the potential barrier, respectively. Note that for $\l \geq
0.026$, $f_0^{\rm eq} = 0.5$ and $\langle X\rangle_{\rm eq} = X_{\rm
max}$.

\vspace{2.cm}

{\bf Figure 4}:  Comparison between volume-averaged field and location of the
inflection point $X_{{\rm inf}}$ and top of the barrier $X_{{\rm max}}$,
as a function of $\l$. \\

\vspace{2.cm}

Recalling the information from Fig. 3, we conclude that there
is a clear distinction between the `smooth' and `mixed' regimes. The
fact that the mean-field potential used in the simulation has a barrier
between the two phases does not imply that we can assume that the
background is sufficiently smooth in order for homogeneous nucleation
to be applicable. Thermal fluctuations, {\it not included} in the
coarse-grained potential, change the character of the transition by
promoting an efficient mixing between the two stable phases. In other
words, the effective potential describing the true behavior of the system
turns into a single-well potential with minimum at $X_{\rm max}$ for
$\lambda > \lambda_c$, even though the mean-field potential has a barrier
between the two phases. Similar behavior exists in ferromagnets, where it is
known that the true Curie temperature is below that predicted by mean-field
theory. In order to improve on the mean-field results, one uses large N or
$\varepsilon$-expansion methods, which sum over a larger class of diagrams
incorporating leading infrared effects. It is possible that our results
could be understood in part by using $\varepsilon$-expansion methods.
However, in the next Section I will propose a different approach to studying
phase mixing, based on a kinetic equation for subcritical
nonperturbative fluctuations,
which, I believe, are responsible for the mixing. This is a new version of
the subcritical bubbles method proposed by
Gleiser, Kolb, and Watkins in Ref. \cite{GKW}. It will be shown that this
method provides a simple and powerful way to study the background.

\section{Modeling Phase Mixing with Subcritical Bubbles}

As was stressed before, the computation of decay rates based on homogeneous
nucleation theory assumes a smooth metastable background over which critical
bubbles of the lower free energy phase will appear, grow and coalesce,
as the phase transition
evolves. However, as the results from the numerical simulation indicate,
the assumption of smoothness is not always valid. To the skeptical reader,
I point out that several condensed matter experiments indicate
that homogeneous nucleation fails to describe the transition
when the nucleation barrier ($\Delta S/T$) becomes too small. Furthermore,
the agreement between theory and experiment has a long and problematic
history \cite{NUCEXP}. Homogeneous nucleation has to be used with care, in
a case by case basis. It is thus important to obtain a criterion which will
allow us to specify the conditions when homogeneous nucleation theory does
apply. Below I describe a simple method which does exactly that.

The basic idea is that in a hot system, not only small but also large
amplitude fluctuations away from equilibrium will, in principle,
be present. Small
amplitude fluctuations are perturbatively
incorporated in the evaluation of the finite
temperature effective potential, following well-known procedures. Large
amplitude fluctuations probing the nonlinearities of the theory are not.
Whenever they are important, the perturbative effective potential becomes
unreliable. In an ideal world, we should be able to sum over all amplitude
fluctuations to obtain the exact partition function of the model, and
thus compute the thermodynamic quantities of interest. However, we can only
to this perturbatively, and will always miss information coming from the
fluctuations not included in its evaluation. If large amplitude fluctuations
are strongly suppressed, they will not contribute to the partition function,
and we are in good shape. But when are they important? We can try to
approach this question avoiding complicated issues related to the
evaluation of path integrals beyond the Gaussian approximation by obtaining
a kinetic equation which describes the fraction of volume populated by these
large amplitude fluctuations. In order to keep the treatment simple, and
thus easy to apply,
several assumptions are made along the way, which I believe are quite sensible.
In any case, the strength of the method is demonstrated when the results are
compared with the numerical experiments described before.

Large amplitude fluctuations away from equilibrium are modelled by
Gaussian-profile spherically-symmetric
field configurations of a given size and amplitude. They can be thought of
as being coreless bubbles. Keeping
with the notation of the numerical experiment,  fluctuations away from the
0-phase, and into the 0-phase are written respectively as,

\beq
\phi_c(r)=\phi_ce^{-r^2/R^2}~,~~~\phi_0(r)=\phi_c\left (1-
e^{-r^2/R^2}\right ) \:,
\eeq
where $R$ is the radial size of the configuration, and $\phi_c$ is the
value of the amplitude at the bubble's core,
away from the 0-phase. In previous treatments
(cf. Refs. \cite {GKW} and \cite{GG}), it was assumed that $\phi_c=\phi_+$,
that is, that the configuration interpolated between the two minima of the
effective potential, and that $R=\xi(T)$, where $\xi(T)=m(T)^{-1}$
is the
mean-field correlation length. But in general, one should sum over all
radii and amplitudes above given
values which depend on the particular model under
study. This will become clear as we go along.

Define $dn(R,\phi,t)$
as the number density of bubbles of radius between $R$ and
$R+dR$ at time $t$, with amplitudes $\phi\geq \phi_c$ between $\phi$ and
$\phi+d\phi$. By choosing to sum over bubbles of amplitudes $\phi_c$
and larger, we are effectively describing the system as a ``two-phase''
system. For example, in the numerical simulation above it was assumed
that the 0-phase was for amplitudes $\phi \leq \phi_{\rm max}$, and the
+-phase was for amplitudes $\phi > \phi_{\rm max}$. Clearly, for a
continuous system this division is artificial. However, since the models
we are interested in have two local minima of the free energy, this
division becomes better justified. Fluctuations with small enough amplitude
about the minima
are already summed over in the computation of the effective potential.
It is the large amplitude ones which are of relevance here. To simplify the
notation, from now on I will denote by ``+-phase''
all fluctuations with amplitudes
$\phi > \phi_c$ and larger. The choice of $\phi_c$ is model-dependent, as will
be clear when we apply this formalism to specific examples.

The fact that the bubbles shrink will be incorporated in the
time dependence for the radius $R$\footnote{Of course, the amplitude $\phi$
will also be time-dependent. However, its time-dependence is coupled to
that of the radius, as recent studies have shown \cite {OSC}.
In order to
describe the effect of shrinking on the
population of bubbles it is sufficient to include only the
time dependence of the radius.}.
Here, I will
only describe a somewhat simplified approach to the dynamics. More details
are provided in the forthcoming paper by Gleiser, Heckler, and Kolb \cite{GHK}.
The results, however, are essentially identical.

The net rate at which bubbles of a given
radius and amplitude are created and destroyed is given by the kinetic
equation,
\begin{eqnarray}
\label{eq:KIN}
{{\partial \n}\over {\partial t}}=-{{\partial \n}\over {\partial R}}
\left ({{dR}\over {dt}}\right )+\left ({{V_0}\over {V_T}}\right )\Gamma_{0
\rightarrow +}(R)  \nonumber\\
 - \left ({{V_+}\over {V_T}}\right )\Gamma_{+\rightarrow 0}(R)
- \left(\frac{V_+}{V_T}\right) \Gamma_{TN} (R)
\end{eqnarray}
Here, $\Gamma_{0\rightarrow +}(R)$ ($\Gamma_{+\rightarrow 0}(R)$)
is the rate per unit volume for
the thermal nucleation of a bubble of radius $R$ of +-phase  within
the 0-phase (0-phase within
the +-phase). $\Gamma_{TN}  (R) \simeq a
T/\frac{4}{3} \pi R^3$ is the (somewhat ad hoc)
expression used for the thermal destruction
rate, with $a$ a constant related to the coupling to the thermal bath.
$V_{0(+)}$ is the volume available for nucleating bubbles of the
+(0)-phase. Thus we can write, for the total volume of the system,
$V_T=V_0+V_+$, expressing the fact that the system has been ``divided'' into
two available phases, related to the local minima of the free energy density.
It is convenient to define the fraction of volume in the +-phase,
$\gamma$, as

\beq
{{V_0}\over {V_T}} \equiv 1 - \gamma ~~.
\eeq

In order to compute $\gamma$ we must sum over {\it all} bubbles of different
sizes, shapes, and amplitudes within the +-phase, {\it i.e.}, starting with
$\p_{\rm min} \geq \p_c$.
Clearly, we cannot compute $\gamma$ exactly. But it turns out
that a very good approximation is obtained by assuming that the bubbles are
spherically symmetric, and with radii above a given minimum radius, $R_{\rm
min}$. The reason we claim that the approximation is good comes from comparing
the results of this analytical approach with numerical simulations.
The approximation starts to break down as the
background becomes more and more mixed, and the morphology of the ``bubbles''
becomes increasingly
more important, as well as other terms in the kinetic equation which
were ignored. For example, there should be a term which accounts for
bubble coalescence, which increases the value of $\gamma$. This term becomes
important when the density of bubbles is high enough for the probability
of two or more of them coalescing to be non-negligible. As we will see,
by this point the mixing is already so pronounced that we are justified in
neglecting this additional complication to the kinetic equation. As a bonus,
we will be able to solve it analytically. The expression for $\gamma$ is,

\beq\label{e:gamma}
\gamma \simeq \int_{\p_{\rm min}}^{\infty}\int_{R_{\rm min}}^{\infty}
\left ({{4\pi R^3}\over 3}\right ){{\partial^2 n}\over {\partial \p\partial R}}
d\p dR~~.
\eeq

The attentive reader must have by now noticed that we have a coupled
system of equations; $\gamma$, which appears in the rate equation for the
number density $n$, depends on $n$ itself. And, to make things even worse,
they both depend on time. Approximations are in order,
if we want to proceed any further along an analytical approach. The first thing
to do is to look for the equilibrium solutions, obtained by setting
$\partial n / \partial t =0$ in the kinetic equation. In equilibrium, $\gamma$
will also be constant\footnote{This doesn't mean that thermal
activity in or between the two
phases is frozen; equilibrium is a statement of the average
distribution of thermodynamical quantities. Locally, bubbles will be
created and destroyed, but always in such a way that the average value of
$n$ and $\gamma$ are constant.}. If wished, after finding the equilibrium
solutions one can find the time-dependent solutions, as was done in
Ref. \cite{GG}. Here, we are only interested in the final equilibrium
distribution of subcritical bubbles, as opposed to the approach to
equilibrium.

The first approximation is to take the
shrinking velocity of the bubbles to be constant, $dR/dt = -v$. This is in
general not the case ({\it cf.} Ref. \cite {OSC}),
but it does encompass the fact that subcritical bubbles
shrink into oblivion.
The strength of the thermodynamic approach is that details of how
the bubbles disappear are unimportant, only the time-scale playing a r\^ole.
The second approximation is to assume that the rates for creation and
destruction of subcritical fluctuations are Boltzmann suppressed, so that
we can write them as $\Gamma= AT^4e^{-F_{\rm sc}/T}$, where $A$ is an arbitrary
constant of order unity, and $F_{\rm sc}(R,\p_c)$
is the cost in free energy to
produce a configuration of given radius $R$ and core amplitude $\p_c$. For the
Gaussian {\it ansatz} we are using, $F_{\rm sc}$ assumes the general
form, $F_{\rm sc}=\alpha R +\beta R^3$, where $\alpha = b\p_c^2$ ($b$ is
a combination of $\pi$'s and other numerical factors)
and $\beta$ depends on
the particular potential used. In practice, the cubic term can usually be
neglected, as the free energy of small ($R\sim \xi$) subcritical bubbles
is dominated by the gradient (linear) term.
We chose to look at the system at the
critical temperature $T_c$. For this temperature, the creation and destruction
rates, $\Gamma_{0\rightarrow +}$ and $\Gamma_{+\rightarrow 0}$ are identical.
Also, for $T_c$, the approximation of neglecting the cubic term is very good
(in fact it is better and better the larger the bubble is)
even for large bubbles, since for degenerate vacua
there is no gain (or loss) of volume energy for
large bubbles. Finally, we use that $V_+/V_T = \gamma$ in the $\Gamma_{+
\rightarrow 0}$ term, and that $V_+/V_T = n(R)V_+(R)$ in the $\Gamma_{\rm TD}$
term. This latter expression assumes that the distribution of bubbles is
sharply peaked around the smallest bubbles considered, so that the sum over
all radii is fairly well approximated by one term. A more sophisticated
approach is presented in Ref. \cite {GHK}.
We can then write the equilibrium rate equation as,

\beq
{{\partial n}\over {\partial R}} = dn(R) - cf(R)~~,
\eeq
where,

\beq
d\equiv aT/v, ~~c\equiv (1-2\gamma)AT^4/v,~~f(R)\equiv e^{-F_{\rm sc}/T}~~.
\eeq

Integrating from $R_{\rm min}$ and imposing that $n(R\rightarrow\infty)=0$,
the solution is easily found to be,

\beq
n(R) = {c\over {\left (d+ \alpha(\p_c)/T \right )}}e^{-\alpha(\p_c) R/T}~~.
\eeq

Not surprisingly, the equilibrium number density of bubbles is Boltzmann
suppressed. But we now must go back to $\gamma$, which is buried in the
definition of $c$. We can solve for $\gamma$ perturbatively, by plugging
the solution for $n$ back into Eq. \ref{e:gamma}. After a couple of
fairly nasty integrals, we obtain,

\beq\label{e:gamma_sol}
\gamma = {{g\left (\alpha(\p_{\rm min}),R_{\rm min}\right ))}\over
{1 + 2 g\left (\alpha(\p_{\rm min}),
R_{\rm min}\right )}}~~,
\eeq
where,

\beq
g\left (\alpha(\p_{\rm min}),R_{\rm min}\right )
 = {{4\pi}\over 3}\left ({{AT^4}\over v}\right )
\left ({T\over {\alpha}}\right )^3{{e^{-\alpha R_{\rm min}/T}}\over {\left
(d+ \alpha/T \right )}}\left [6+\left ({{\alpha R_{\rm min}}\over T}\right )^3
+3{{\alpha R_{\rm min}}\over T}\left (2+{{\alpha R_{\rm min}}\over T}\right )
\right ]~~.
\eeq

\subsection{Comparison with the numerical simulations}

We can now apply this formalism to any model we wish. The first obvious
application is to compare $\gamma$ obtained from the numerical experiment
with the value obtained from the kinetic approach. From the definition
of the equilibrium fractional population difference, Eq. \ref{e:Delta},

\beq
\Delta F_{\rm EQ}\left (\theta_c \right ) = 1 -2\gamma~~.
\eeq

Thus, it is straightforward to extract the value of $\gamma$
from the numerical simulations, as a function of $\l$.
Also, as we neglected the volume contribution to the free energy of
subcritical bubbles, we have,

\beq
F_{sc} = \alpha(\p_c)R_{\rm min} = {{3\sqrt{2}}\over 8}\pi^{3/2}X_-^2
(\theta_c)R_{\rm min} ~~,
\eeq
where, as you recall, $X_-$ is the position of the maximum of the mean-field
potential used in the simulations. So, we must sum over all amplitudes
with $X\geq X_-$, and all radii with $ R\geq 1$ (in dimensionless units),
as we took the
lattice spacing to be $\ell=1$. That is,
we sum over all possible sizes, down to
the minimum cut-off size of the lattice used in the simulations. In practice,
we simply substitute $\p_c=X_-$ and $R_{\rm min}=1$ in the expression
for
$\gamma$.
In Fig. 5, we compare the numerical
results for $\gamma$ (dots) with the results from the analytical integration
of the kinetic equation. We took $A/v=1$, and the two curves are for
$a/v=0$ (no thermal destruction), and $a/v=1$. Up to the critical
value for $\l\simeq 0.025$, the agreement is very convincing. As we increase
$\l$ into the mixed phase region of the diagram, the kinetic approach
underestimates the amount of volume in the +-phase. This is not surprising,
since for these values of $\l$
the density of subcritical bubbles is high enough that
terms not included in the equation become important, as I mentioned before.
However, the lack of agreement for higher values of $\l$ is irrelevant,
if we are interested in having a measure of the
smoothness of the background; clearly, the rise in $\gamma$ is sharp enough
that homogeneous nucleation should not be trusted for $\l > 0.024$ or so,
as the fraction of volume occupied by the +-phase is already around 30\%
of the total volume. Subcritical bubbles give a simple and quantitatively
accurate picture
of the degree of inhomogeneity of the background, offering a guideline as to
when homogeneous nucleation theory can be applied with confidence.

\vspace{2.cm}
{\bf Figure 5}:  Comparison between results from numerical simulations
(dots) and the result from the kinetic equation. \\

\vspace{2.cm}

\subsection{ Thermal Phase Mixing in the Standard Electroweak Model}

It is straightforward to compute $\gamma$ for the electroweak model.
One can choose any effective potential, and simply plug the results
into Eq. \ref{e:gamma_sol}. Here, I will work with the simplest version
of the model, the unimproved 1-loop potential, written
in Eq. \ref{e:VEW}. Clearly, different effective potentials will give
different values for $\gamma$.
So, if we used instead the 2-loop potential
of Ref. \cite{2-loop}, we would obtain smaller values for $\gamma$, as
the two-loop potential gives rise to a stronger transition. The point here
is not to get into the problems of computing a reliable effective potential,
but to apply the method just developed to a simple situation.

There have been several recent
papers on applying subcritical bubbles to different effective potentials
or exploring some of their properties\cite{SUBNEW}.
In my view, there are two important points which seem to be overlooked
in most (but not all) of these
works. First, the importance of subcritical bubbles is sensitive
to the parameters of the model. If one finds that subcritical bubbles
are not relevant for a given range of parameters, this does not mean
that they are ruled out as unphysical. It simply means that they
will be relevant for a different range (usually higher Higgs masses
if one is interested in the electroweak phase transition);
hot systems fluctuate, even if our description of the physics through a
perturbative effective potential may be  inaccurate. For the particular
case of the electroweak transition, we eagerly await for lattice gauge
simulations which will probe the regime of higher Higgs masses.
Second, approaches that use a saddle-point evaluation of the
partition function to compute, say, the dispersion of the field around
the background, will miss the important contributions coming from
large amplitude fluctuations. Thus, they will underestimate the amount
of mixing. This is an advantage of the kinetic approach presented here.

Using the parameters for the electroweak model, we can express
$\gamma$ in terms of the Higgs mass,
as shown in Fig. 6. I took $R_{\rm min}=\xi(T_c)$, and $\p_c=\p_+(T_c)$. Again,
the results are sensitive to the choice of minimum radius and amplitude,
but these values are certainly conservative. It is clear from the Figure
that taking $v=1$, $\gamma$ changes sharply from $0$ to $0.5$ for
$60<{\rm M_{\rm Higgs}}<70$, the window where lies
the present lower bound on the
Higgs mass. Thus, these results appear to rule out nucleation for the
unimproved 1-loop effective potential\footnote{There is a volume factor
which suppresses $\gamma$ not included here; simply, not all of the
subcritical bubble is in the +-phase, just a fraction of it. However,
this suppression is compensated by taking the more realistic choice
$\p_c=\p_{\rm max}$, as opposed to the
conservative choice $\p_c=\p_+$ we used.}.
Mechanisms for baryogenesis
based on homogeneous nucleation must take the mixing of the background
due to subcritical fluctuations into account.

\vspace{2.cm}

{\bf Figure 6}:  Thermal phase mixing for the minimal electroweak model
 as a function of the Higgs mass.\\

\vspace{2.cm}

\section{Concluding Remarks}

In this talk I presented the results of both numerical simulations and
analytical modelling of phase mixing induced by thermal fluctuations. Although
I stressed mostly the importance of these results in the context of homogeneous
nucleation in finite temperature field theory,
the ideas presented here can be adapted to a variety of
different situations of interest also in laboratory applications, as long as
the system in question can be described by a coarse-grained Ginzburg-Landau
free energy with a non-conserved order parameter. For example, together with
Andrew Heckler, a method to compute the influence of subcritical bubbles on
the nucleation rate is presently
being developed, which, we believe, is quite general
\cite{GH}. The basic idea is to include the available free energy in the gas of
subcritical bubbles in the computation of the effective nucleation barrier.
In practice, this is done by obtaining a new coarse-grained effective action
which sums over fluctuations of subcritical size, in the spirit of the
renormalization group. Comparing the results from our method to numerical
simulations of nucleation in 2 dimensions (Ref. \cite{AG}) shows how
subcritical fluctuations can account for a dramatic drop in the effective
nucleation barrier which is observed close to degeneracy.

I hope it was clear from this talk that subcritical bubbles can provide
a clear and quantitatively accurate description of the degree of homogeneity
of a given background, metastable or not. It would be interesting to
generalize the ideas presented here to quantum fluctuations as well, since
we should expect them to be dominant at very low temperatures. Naively,
we could think of quantum subcritical bubbles as nonperturbative
violations of energy
conservation, with a probability distribution
suppressed by their Euclidean action. If present, these fluctuations could
have many interesting consequences, from inflationary scenarios to
nucleation of topological defects in systems with nontrivial vacuum
topology.

\vspace{2.cm}

\noindent{\bf Acknowledgements}
\vspace{1.cm}

I am grateful to my collaborators Julian Borrill, Andrew Heckler and
Rocky Kolb, as well
as to M. Alford, E. Copeland, H.-R. M\"uller, and
R. Ramos for the many long discussions on bubbles and
phase transitions.
I am also grateful to the
Nasa/Fermilab Astrophysics Center for their kind hospitality where some of the
work presented here was developed.

At Fermilab I acknowledge partial support from
NASA Grant NAG 5-2788 and DOE.
This work was partially supported at Dartmouth
by the National Science Foundation through a
Presidential
Faculty Fellows Award no. PHY-9453431 and by Grant no. PHYS-9204726, and
by a NASA Grant NAGW-4270.

Finally, I thank  Hector de Vega and Norma Sanchez
for their warm hospitality during this workshop.

\vspace{1.5cm}


\vfill\eject

\centerline {\bf TABLE 1}
\vspace{2.in}

\begin{center}
\begin{tabular}{|c|c|c|}\hline\hline
$\l $ & $\tau_{{\rm EQ}}$ & $\sigma $ \\
\hline\hline
 0.015 & 25.0  & 1.0 \\
 0.020 & 45.0  & 1.0 \\
 0.022 & 60.0  & 1.0  \\
 0.024 & 110.0 & 0.80   \\
 0.026 & 220.0 & 0.50 \\
 0.028 & 120.0 & 0.75 \\
 0.030 & 100.0 & 0.80 \\
 0.035 & 55.0  & 0.80  \\
 0.040 & 40.0  & 0.90 \\
\hline
\end{tabular}
\end{center}

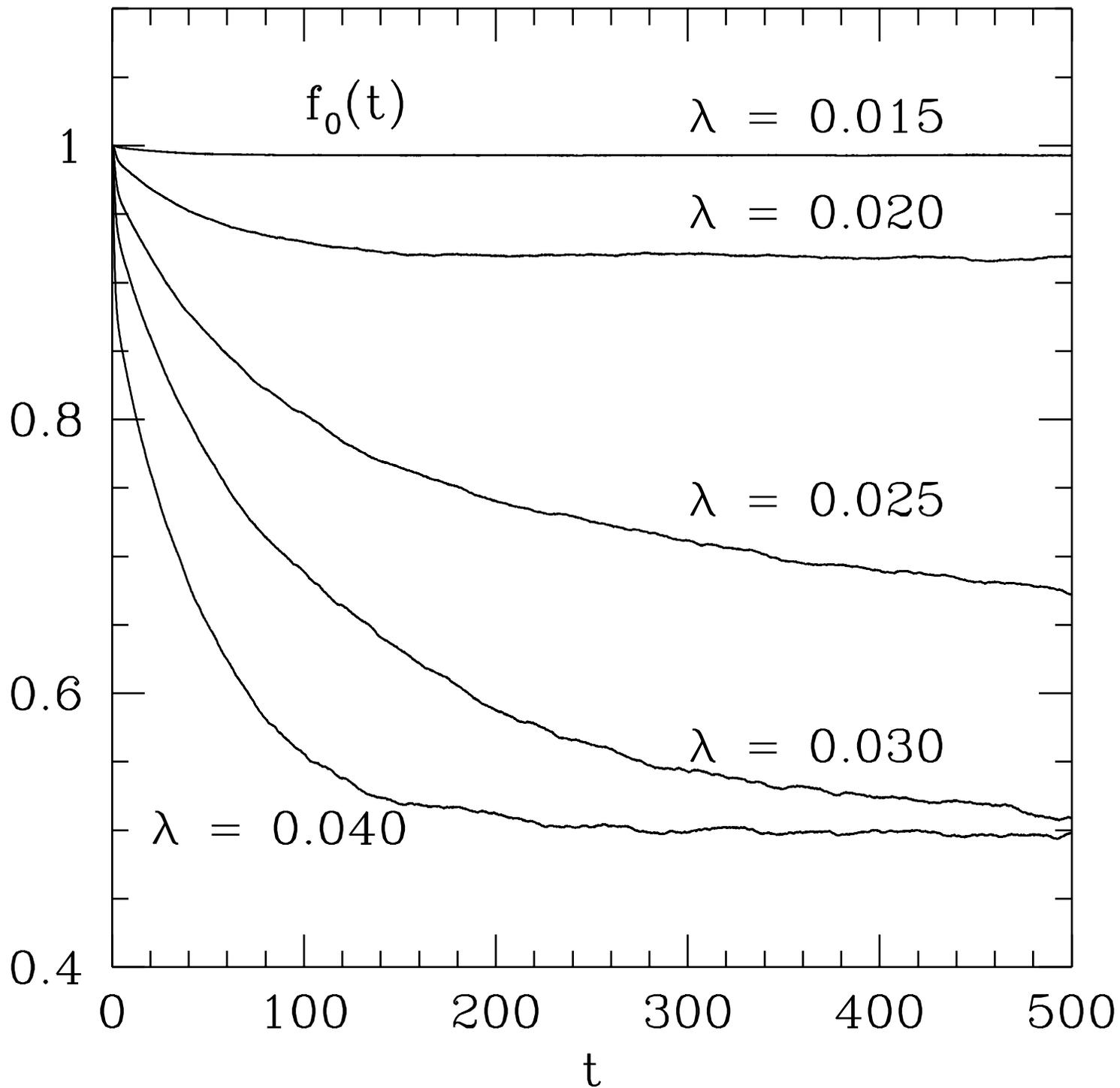
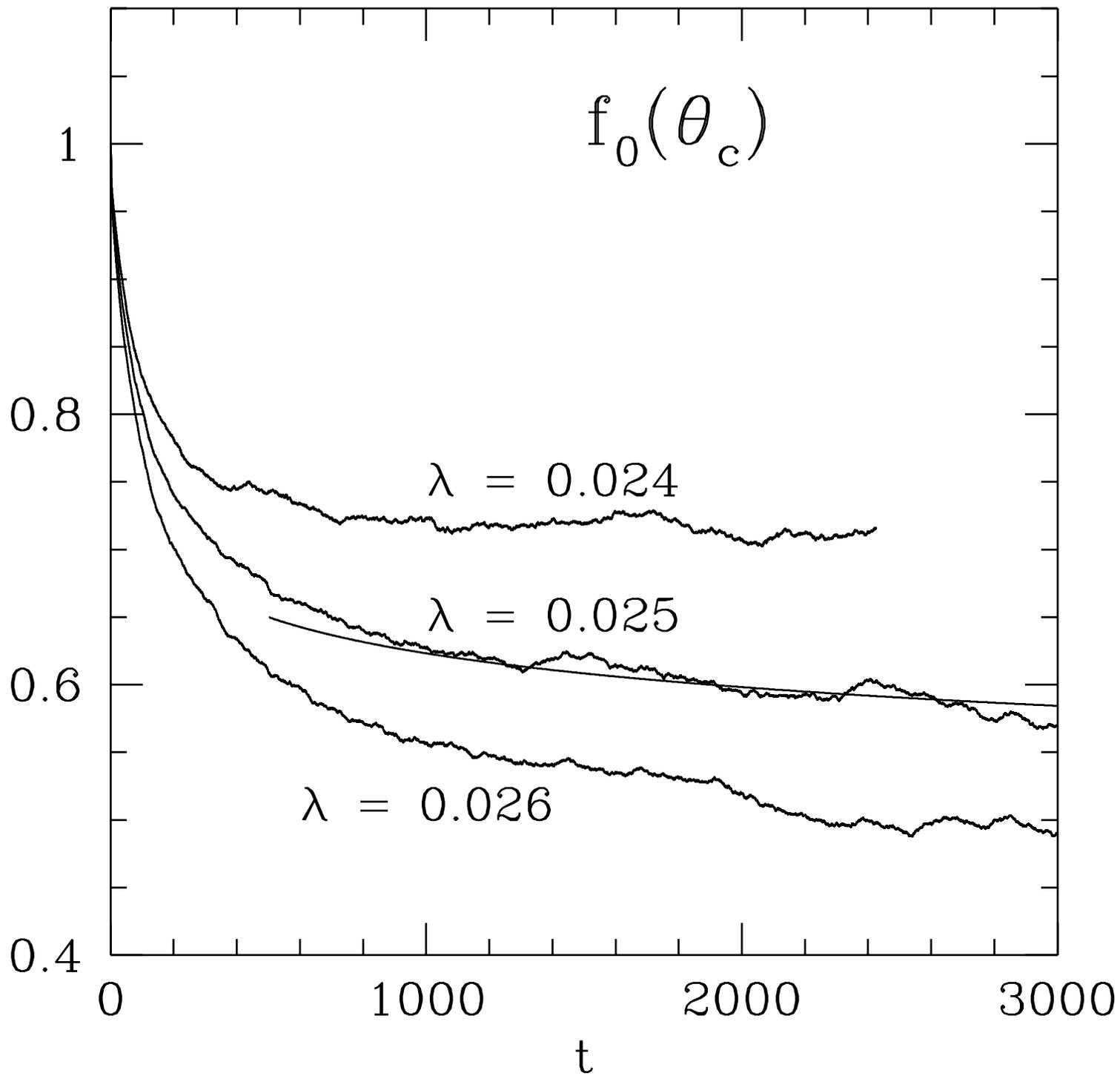
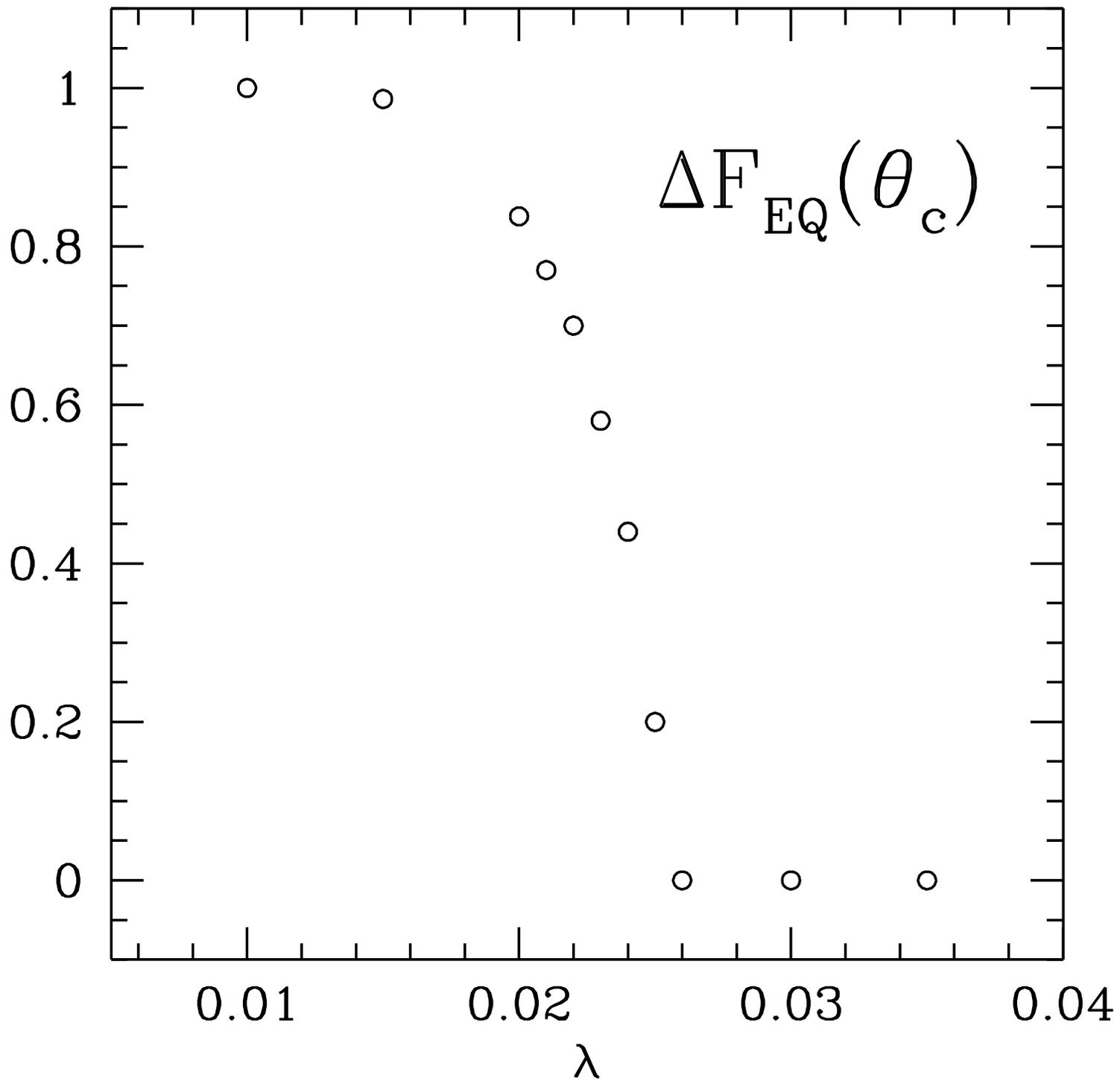
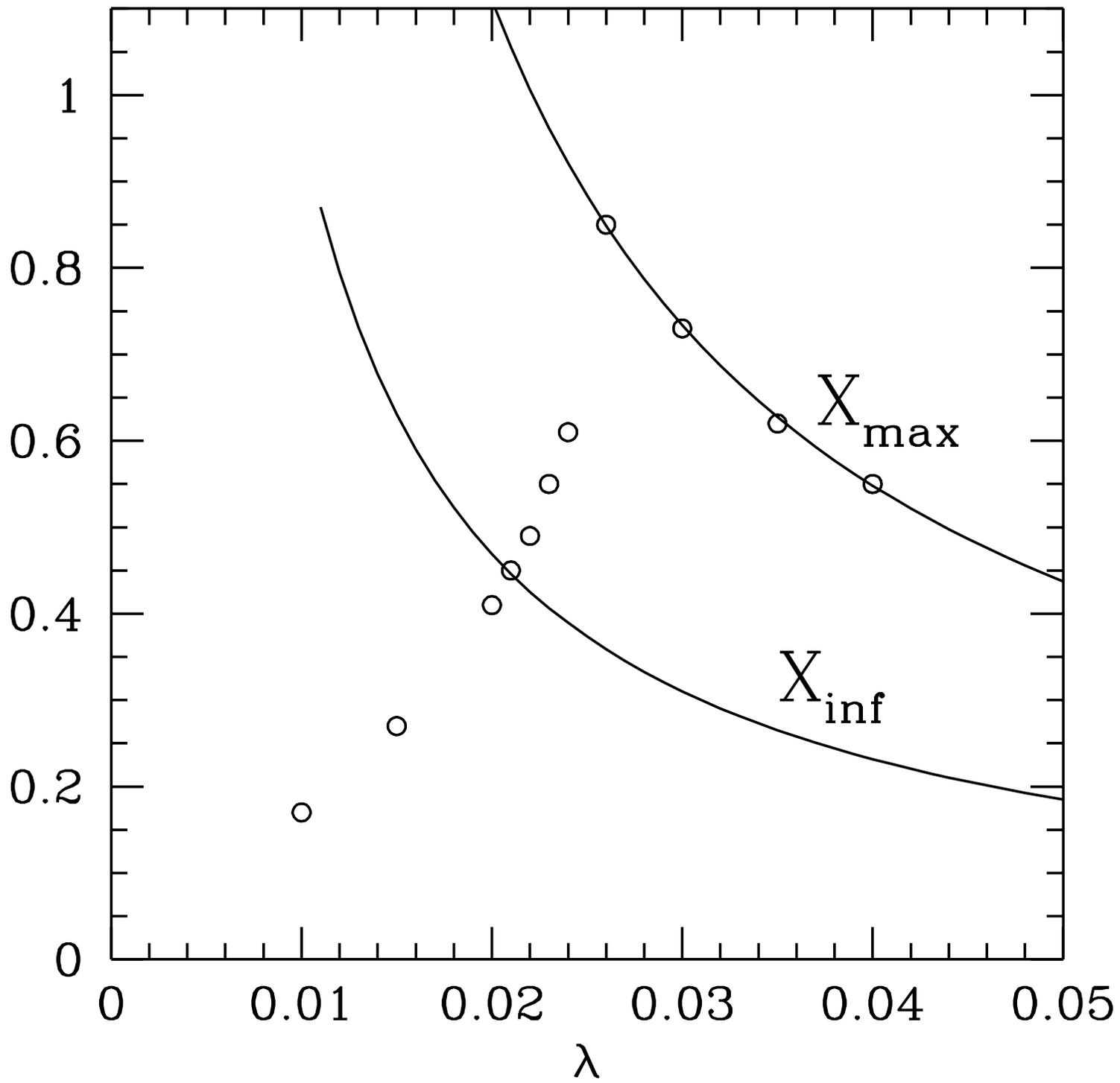
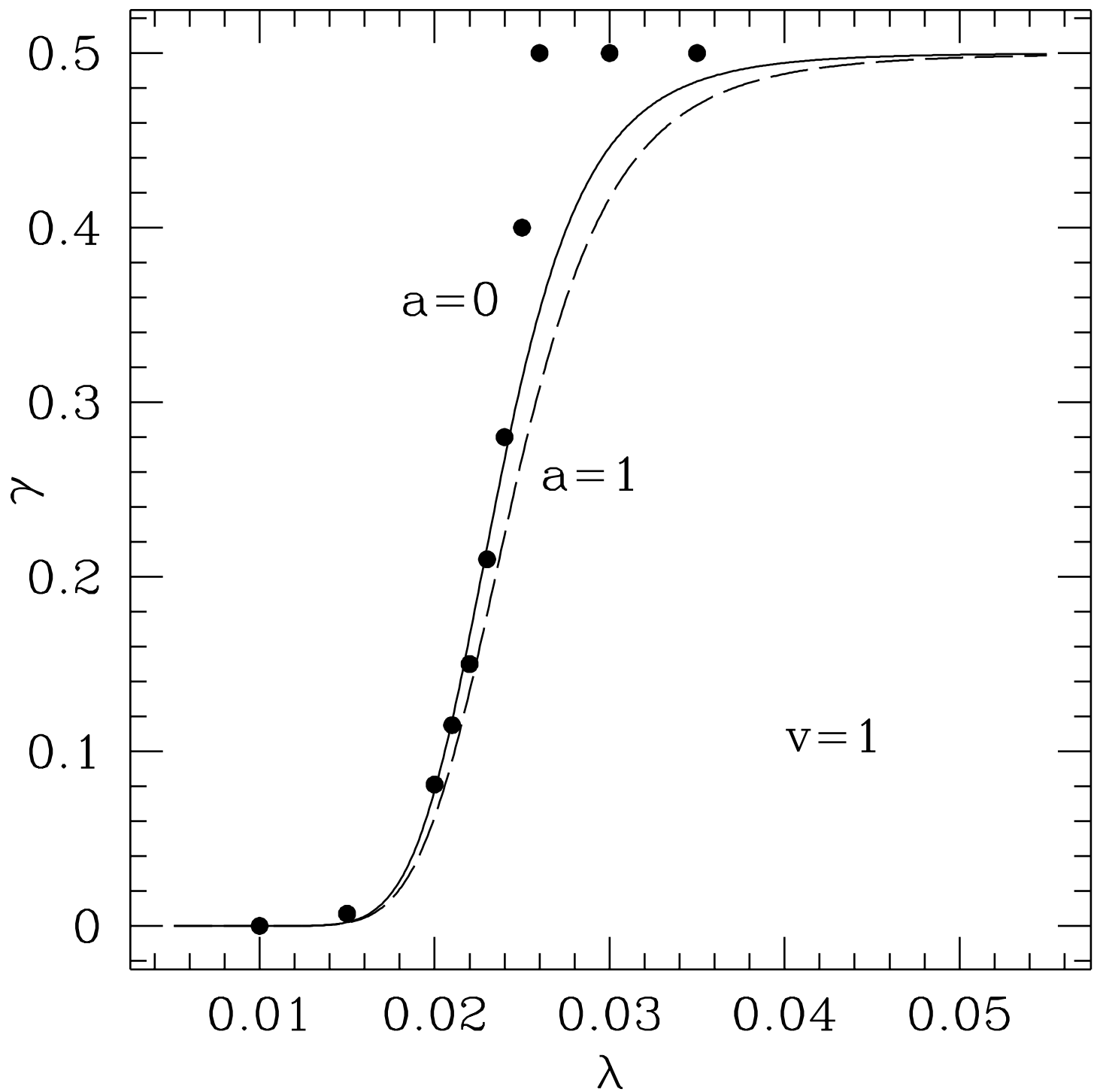
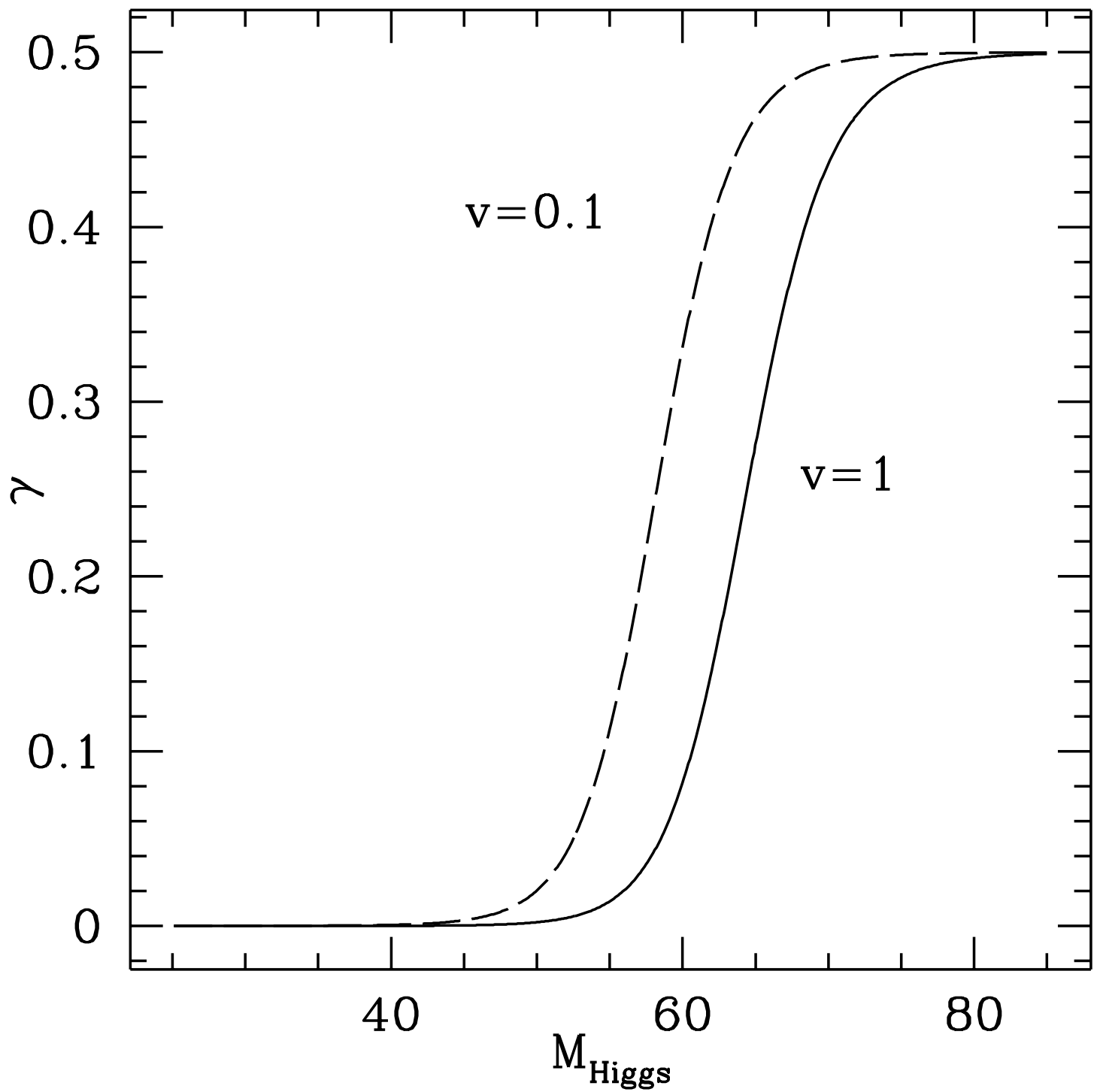
\end{document}